\pdfoutput=1
\documentclass[prb,aps,superscriptaddress,showpacs,twocolumn,amssymb,amsmath,amstext,amsfont,noshowkeys]{revtex4}

\usepackage{bm}
\usepackage{graphicx}
\usepackage{hyperref}

\begin{document}

\title{Excited electron-bubble states in superfluid $^4$He:  a time-dependent density functional approach}

\author{David Mateo}
\affiliation{Departament ECM, Facultat de F\'{\i}sica, and IN$^2$UB,
Universitat de Barcelona. Diagonal 647, 08028 Barcelona, Spain}

\author{Dafei Jin}
\affiliation{Department of Physics, Brown University, Providence, RI
02912, USA}

\author{Manuel Barranco}
\affiliation{Departament ECM, Facultat de F\'{\i}sica, and IN$^2$UB,
Universitat de Barcelona. Diagonal 647, 08028 Barcelona, Spain}

\author{Mart\'{\i} Pi}
\affiliation{Departament ECM, Facultat de F\'{\i}sica, and IN$^2$UB,
Universitat de Barcelona. Diagonal 647, 08028 Barcelona, Spain}

\begin{abstract}

We present a systematic study on the excited electron-bubble states in
superfluid $^4$He using a time-dependent density functional approach. For the
evolution of the 1P bubble state, two different functionals accompanied
with two different time-development schemes are used, namely an accurate
finite-range functional for helium with an adiabatic approximation for electron
versus an efficient zero-range functional for helium with a real-time evolution
for electron. We make a detailed comparison between the quantitative results
obtained from the two methods, which allows us to employ with confidence the optimal
method for suitable problems. Based on this knowledge, we use the finite-range
functional to calculate the time-resolved absorption spectrum of the 1P bubble, 
which in principle can be experimentally determined, and we use the
zero-range functional
to real-time evolve the 2P bubble for several hundreds of picoseconds, which 
is theoretically interesting due to the break down of adiabaticity for this state. 
Our results discard the physical realization of relaxed, metastable 2P electron-bubbles.



\end{abstract}

\date{\today}

\maketitle

\section{Introduction}

Electron bubble (e-bubble) in liquid-helium has been an attractive topic for
numerous experimental and theoretical studies in the past, and it has
drawn again some interest in recent years.
\cite{Elo02,Gho05,Gra06,Ros06,Pi07,Leh07,Mar08,Guo09,Anc09,Mat10,Jin10} Density
functional (DF) theory has proved to be a powerful tool in dealing with many
interesting physical situations involving electron bubbles.
When it is applied to optically excited
e-bubbles, not only can it achieve quantitative agreement with experiments on
the absorption spectra,\cite{Elo02,Gra06,Gri90,Gri92} it can also
nicely display the dynamical evolutions on the picosecond time scale, such as how the
bubbles change shapes, release energy, or even break into smaller
bubbles.\cite{Jin10a,Mat10a} These latest works likely require using
different density functionals (finite-range or zero-range for liquid-helium) in
different time-development schemes (adiabatic or real-time evolutions for
electron). Regardless of the technical details, the time-dependent density
functional approach is no doubt the only workable approach at present for
studying the evolution of the excited e-bubble states in liquid-helium.
The quantitative
results drawn from these simulations can be useful to interpret the
experimental results and predict new ones.

Upon dipole excitation from the 1S ground state to the 1P or 2P
excited state, an e-bubble evolves by relaxing its shape around the
excited electron probability density. This relaxation  eventually drives
the e-bubble back to the spherical 1S ground state. It has been quantitatively
shown\cite{Mat10a} that depending on the pressure $(P)$ applied to
the liquid, this may happen in two different ways. At a pressure below about 1
bar, the e-bubble undergoes damped oscillations for a period of time long
enough to allow the electron to radiatively decay to the deformed 1S state,
which then evolves radiationlessly to the spherical 1S state. In contrast, at 1
bar or above, the excited e-bubble evolves towards a configuration made of two
baby bubbles, so that the probability of finding the electron evenly
distributes between them. This two-bubble configuration is
unstable against asymmetric perturbations, and one expects this instability to
cause the electron to localize in one of the baby bubbles while the other
collapses.\cite{Mat10a,Mar04,Gri90}

When doing this calculation, the authors of Ref.~\onlinecite{Mat10a} naturally
chose the so-called Orsay-Trento DF.\cite{Dal95} This functional is
finite-range and incorporates a term that mimicks back-flow effects in order
to accurately reproduce the dispersion relation of the elementary
excitations of superfluid $^4$He. 
This is instrumental to properly describe the energy transfer from
the bubble to the liquid, that proceeds by causing all sorts of possible
excitations in the superfluid. The token one has to pay for its use is
the very
high computational cost. Due to the large difference between the intrinsic time
scales of electron and helium, this functional is not well adapted for
fully real-time, three-dimensional evolutions. Considering this
limitation, the adiabatic approximation was used to update the electron 
wavefunction at every instantaneous helium configuration.\cite{Mat10a}
One of the main concerns is then to establish how long the adiabatic
approximation is valid for. A careful analysis led to the conclusion that 
it holds for the 1P e-bubble for at least several tens of
picoseconds,\cite{Mat10a} a period of time large enough to guarantee 
the reliability of the results obtained for this state.
Contrarily, the adiabatic approximation breaks down very quickly for the
2P e-bubble, implying that the existence of relaxed
quasi-equilibrium 2P bubbles is questionable.

The authors of Ref.~\onlinecite{Jin10a} followed a different path. Instead of
using a finite-range DF, they employed a much simpler zero-range
one. This allowed them to carry out fully real-time
calculations for
hundreds of picoseconds without imposing any adiabatic assumption, at the price
of an inaccurate description of the elementary excitations of the
liquid.
Although this may not qualitatively affect the physical results, it is
unclear how quantitatively reliable are the results so obtained.

In this work, we
perform a more systematic study on the relaxation of the excited e-bubbles. We
first carry out a detailed comparison between the finite-range and zero-range
functionals applied to the 1P bubble problem, through which we gain some
insight about their strongpoints and shortcomings. We then use the
finite-range functional to calculate the time-resolved absorption
spectrum of the 1P bubble, which can in principle be measured in the
experiments. Next, we use
the zero-range functional to real-time evolve the 2P bubble for several
hundreds of picoseconds, which is of theoretical interest as one can clearly
trace how the adiabatic approximation breaks down for this state.

Our paper is organized as follows. In Sec. II, we introduce the theoretical
framework and numerical schemes. In Sec. III, we discuss the
results so obtained. Finally, a summary is presented in Sec.
IV. Several movies showing the dynamical evolution of electron bubbles
can be found in the supplementary material.\cite{web}

\section{Theoretical framework}

Within the DF approach,
the energy of an electron-helium system at zero temperature can be written as
a functional of the single-electron wavefunction $\Phi$ and the macroscopic
helium wavefunction $\Psi$

\begin{equation}
\begin{split}
E[\Phi,\Psi] = &\, \frac{\hbar^2}{2 m_{\mathrm{e}}} \int d \bm{r}\,
|\nabla\Phi|^2 + \frac{\hbar^2}{2 m_{\mathrm{He}}} \int d \bm{r}\,
|\nabla\Psi|^2 \\
& + \int d \bm{r}\, |\Phi|^2\, V_{\mathrm{e-He}}[\rho] + \int d \bm{r}\,
\mathcal{E}_{\mathrm{He-He}}[\rho]\; . \label{eq1}
\end{split}
\end{equation}
Specifically, $\Psi = \sqrt{\rho} \exp [\imath S ]$ gives the helium particle
density $\rho$ and the superfluid velocity $\mathbf{v} = \hbar \nabla S /
m_{\mathrm{He}}$.\cite{Mat10a} $V_{\mathrm{e-He}}[\rho]$ is the electron-helium
interaction potential,\cite{Che94} and
$\mathcal{E}_{\mathrm{He-He}}[\rho]$ is the helium-helium potential energy
density. For the sake of comparison, we choose for $\mathcal{E}_{\mathrm{He-He}}[\rho]$
either the finite-range Orsay-Trento (OT) density functional,\cite{Dal95} or
the zero-range Stringari-Treiner (ST) density functional.\cite{Str87} It is
known that the former one provides a very accurate description of
superfluid
$^4$He, particularly of the dispersion relation that covers all the
roton
excitations up to the wave number $q=2.3$~\AA{}$^{-1}$. In contrast, the
latter
one only reproduces the long-wavelength phonon excitations and is not so
accurate, but it has the advantage of high computational efficiency in
dynamic  
evolutions. The dispersion relations obtained from both functionals 
using the method of Ref.~\onlinecite{Mat10c} are plotted in
Fig.~\ref{fig1} together with the experimental results.\cite{Don81}

Functional variations of the associated grand potential with respect to $\Psi$
and $\Phi$ yield the following Euler-Lagrange equation for the helium and 
Schr\"odinger equation for the electron:

\begin{equation}
-\frac{\hbar^2}{2 m_{\mathrm{He}}}\Delta \Psi + \left\{\frac{\delta
\mathcal{E}_{\mathrm{He-He}}[\rho]}{\delta \rho} +|\Phi|^2\, \frac{\delta
V_{\mathrm{e-He}}[\rho]}{\delta \rho} \right\} \Psi = \mu \Psi
\label{eq2}
\end{equation}
\begin{equation}
-\frac{\hbar^2}{2 m_{\mathrm{e}}}\Delta \Phi + V_{\mathrm{e-He}}[\rho] \Phi  =
\varepsilon \Phi\; , 
\label{eq3}
\end{equation}
where $\mu$ is the helium chemical potential and $\varepsilon$ is the electron
eigenenergy. Throughout this paper, we treat pressure as a given
external condition. The associated chemical potential and particle density 
in bulk liquid are obtained from the equation of state derived from the
DF being used.

The above equations are solved numerically with 13-point finite-difference
formulae. In the finite-range DF case, we
work in three-dimensional Cartesian coordinates that allow for an extensive use
of fast Fourier transformation techniques\cite{FFT} as explained in
Ref.~\onlinecite{Pi07}. Whenever necessary, we implement a Gram-Schmidt scheme
to determine from Eq.~(\ref{eq3}) the electron spectrum in the helium cavity.
In the zero-range DF case, we work in 
cylindrical coordinates assuming azimuthal symmetry around the $z$-axis $(r=0)$
and specular symmetry about the $z=0$ plane. Hence, we only need
to solve the equations in the $r \geq 0$ and $z \geq 0$ ``quadrant'', which
greatly speeds up the calculation and so we can attain longer-time physics. There 
is no difficulty to relax these symmetry restrictions except slowing down the
calculation. However, the results are not expected to be very different since there
is no detectable symmetry-breaking instability in our problems. For both
functionals we use a fairly large spatial step of about 1~\AA{}, without any
apparent loss of numerical accuracy.\cite{Mat10a}

The time evolution starts from an excited e-bubble state, which means that the
electron has been suddenly brought from the 1S onto the 1P or 2P state of the original
spherical e-bubble. From this initial configuration, the superfluid
helium then evolves according to

\begin{equation}
\begin{split}
\frac{\partial\Psi}{\partial t} =& -\frac{\imath}{\hbar} \left\{
-\frac{\hbar^2}{2 m_{\mathrm{He}}}\Delta - \mu + \mathcal{U}_{\mathrm{He-He}}[\rho, \mathbf{v}]\right. \\
& \left. + |\Phi|^2\, \frac{\delta V_{\mathrm{e-He}}[\rho]}{\delta \rho}
\right\} \Psi\; , 
\label{eq4}
\end{split}
\end{equation}
where the detailed form of the effective potential
$\mathcal{U}_{\mathrm{He-He}}[\rho, \mathbf{v}]$ can be found e.g. in Refs.~
\onlinecite{Gia03} and \onlinecite{Leh04}.

In the adiabatic approximation scheme, we do not evolve the electron in real-time
but keep tracing the instantaneous eigenstates satisfying Eq. (\ref{eq3}). 
In most cases, e-bubbles around excited electron states evolve towards
configurations that are
not spherically but axially symmetric. The originally degenerate angular
momentum electron eigenstates in the spherical bubble now split according
to the projection ($m$-values) on the symmetry $z$-axis, among which only the
$\pm m$ states are still degenerate with each other. It is thus convenient to use
the notation for the orbital angular momentum of single particle states in
linear molecules, i.e., $\sigma, \pi, \delta, \phi, \ldots$ for $|m| = 0, 1, 2,
3, \ldots$ In addition to the axial symmetry, 
a $n$P bubble also keeps the original specular symmetry in the
course of its evolution.
Hence, one can construct the electron eigenbasis in such a way that
the electron wavefunctions satisfy 
$\Phi(r,z,\theta)=\pm\Phi(r,-z,\theta)$. The correspondence between
the lower lying spherically and axially symmetric electron states 
is displayed in Fig. \ref{fig2} along with a representation of their
probability densities. The
superscript $+\,(-)$ denotes specularly symmetric (antisymmetric)
states. The adiabatic evolution is obtained by keeping
the 1P (2P) electron in the instantaneous $1\sigma^-$ 
($3\sigma^-$) eigenstate.

In the real-time dynamics scheme, the electron evolves according to
\begin{equation}
\frac{\partial \Phi }{\partial t} = -\frac{\imath}{\hbar}
\left\{-\frac{\hbar^2}{2m_{\mathrm{e}}}\Delta + V_{\mathrm{e-He}}[\rho]\right\}
\Phi\; . 
\label{eq5}
\end{equation}
We employ a fourth-order Runge-Kutta method to obtain the first time steps,
and Hamming's method\cite{Ral60} for subsequent steps. A time step of $10^{-2}$ ps 
is chosen for the adiabatic evolution, and of $10^{-6}$ ps for
the dynamical evolution. These very different values reflect the mass ratio
$m_{\mathrm{e}}/m_{\mathrm{He}}\sim10^{-4}$. This is this reason that
makes a
dynamical evolution unaffordable when $\mathcal{E}_{\mathrm{He-He}}[\rho]$ is
finite-range, since updating the mean field is computationally very
costly.

During the bubble evolution, sound waves released from its surface eventually
reach the cell boundary. If no action is taken, they will bounce
back spoiling the calculation.
A way to handle this problem is to include some source of damping
into Eq. (\ref{eq4}) governing the fluid evolution, see e.g. Refs.
\onlinecite{Elo02,Cer85,Neu08}. We have opted by making  the  
replacement $\imath \longrightarrow \imath + \Lambda(\bm{r})$
in Eq. (\ref{eq4}).
This corresponds to a rotation of time axis in the complex plane by introducing
a damping field $\Lambda(\bm{r})$, which takes the form\cite{Jin10b}

\begin{equation}
\Lambda(\bm{r})= \Lambda_0 \left[ 1 + {\rm tanh}
\left(\frac{s-s_0}{a}\right)\right], \quad s\equiv|\bm{r}| \; .
\label{eq7}
\end{equation}
We keep the dimensionless parameter $\Lambda_0\simeq1.6$, and set $a=5$~\AA{},
$s_0=60$~\AA{} in the finite-range calculation, and $a=8$~\AA{}, $s_0=90$~\AA{}
in the zero-range calculation. The evolution is damping-free
$[\Lambda(\bm{r})\ll1]$ in a sphere of radius $s<s_0-2a$, which is  50~\AA{}
in the finite-range case and 70~\AA{} in the zero-range case. From
Figs.~\ref{fig3}-\ref{fig6}  and the supplementary material\cite{web} one can see that
this region is large enough for the
1P e-bubble to expand within an undampening environment. For the 2P e-bubble
evolution, we use a $(r,z)$ calculation box of $150\times 150$~\AA{}$^2$ and
$s_0=120$~\AA{}, leaving $\sim 100$~\AA{} of undampening space for the bubble to
expand. 

The above  prescription works extremely well, as it efficiently dampens the
excitations of the macroscopic wavefunction at the cell boundaries, and does 
not need a large buffer region to absorb the waves ---actually we use the same 
box where the starting static calculations have been carried out.\cite{Jin10a} 
It allows us to extend the adiabatic calculations of Ref.~\onlinecite{Mat10a} 
from tens to hundreds of picoseconds.

\section{Results and Discussion}

\subsection{1P e-bubble dynamics}

\subsubsection{Adiabatic versus real-time dynamical evolution}

To some extent, an e-bubble in liquid-helium is nearly a textbook example of an
electron confined in a spherical-square-well potential. Its static properties
are fairly insensitive to the complexities of the chosen  
functional provided the bulk and surface properties can be well reproduced.\cite{Gri90,Gri92}
In particular, a zero-range DF description of the 1S-1P
absorption energies of the e-bubble as a function of $P$ does not differ
much from that obtained by a finite-range DF description.\cite{Gra06}
This means that in some situations one may simply use a
zero-range functional,
which has an advantage of high computational speed.

Clearly, the dynamics of an e-bubble is much more
involved than its statics. In our problems, the --nonspherical-- squeezing and
stretching of the bubble may cause its waist to shrink to a point when electron
tunneling plays a role, and may also dissipate a large amount of energy by
exciting elementary modes in the surrounding liquid. So, even if the static
properties of the e-bubble are equally well described by both
functionals, it is not obvious whether they
yield a similar dynamical evolution for an excited e-bubble.
This is the first issue we want to address.

We use two different schemes to compute the relaxation of a 1P e-bubble
at $P=0$, 0.5, 1 and 5~bars, and compare the results so obtained. One such
scheme is the finite-range OT density functional description for helium 
with the adiabatic evolution for electron. The other scheme is the zero-range
ST density functional description for helium with real-time evolution for
the electron.

As can be seen in Figs.~\ref{fig3}-\ref{fig6}, the evolution starts
with the bubble stretching along the $z$-axis and shrinking on its
waist.  After this stage, the bubble may
continue oscillating and releasing energy into the liquid,
eventually reaching a relaxed, metastable 1P state, or may split into 
two baby bubbles due to the liquid filling-in around the bubble waist.
The density waves radiated to the liquid during this evolution take away 
a considerable part of the energy injected into the system during the 
absorption process, i.e., 105 meV at $P=0$ and 148 meV at 5 bar.\cite{Gra06}

In Figs. \ref{fig3}-\ref{fig6} we compare the bubble evolution obtained
within the
two frameworks for different pressures. At a first glance, both dynamics are 
nearly equivalent during the first 50 ps, starting to differ from this time 
on although they are still qualitatively similar.
A more detailed analysis, focused on three key elements of the density
profiles, indicates the following:

\textit{a}. The shape of the bubble surface, defined as the locus where 
the liquid density equals half the saturation density value $\rho_0$,
e.g., 0.0218 \AA$^{-3}$ at $P=0$ bar.
This shape determines the most crucial properties of an e-bubble. 
From this shape, we know whether the bubble is simply connected
or has split. Up to $t \lesssim 50$ ps, the shape of the bubble
is nearly identical in both descriptions. At later times, the
bubble shape changes at a slower pace in the ST than in the OT
description. 

Figure \ref{fig7} illustrates the time evolution of the bubble surface.
In particular, the top panel shows the evolution of
the point on the bubble surface at $r=0$ with $z > 0$. This represents half
the longitudinal extent of the e-bubble.
One can see that this length oscillates
in the ST calculation with a lower frequency than in the OT one. 
If the bubble symmetrically splits into two baby bubbles, there are two
such points, as can be seen in the bottom panel for $P=1$ bar.
We have found that, besides the moment at which the distance between
the baby bubbles increases steadily ---about 175 ps for the OT
functional and 200 ps for ST functional--- there is a time interval
between $\sim$ 60 and $\sim$ 90 ps for the OT functional, and
$\sim$ 60 and $\sim$ 110 for the ST
functional, where the 1P bubble at $P=1$ bar has split but
the emerging baby bubbles are ``locked'' by the shared electron
that exerts some attractive force on them, forcing them back to a
simply connected configuration. Eventually, the baby bubbles
are unlocked and the distance between them grows.

\textit{b}. The surface thickness of the bubble, defined as the width of the
region satisfying $0.1\rho_0 \leq \rho(\bm{r}) \leq 0.9\rho_0$. The thickness 
of the bubble surface has been found to be nearly independent of the local surface 
curvature at anytime during the evolution, see also Ref. \onlinecite{Mat10a}.
It is about 1 \AA{} larger in the ST than in the OT
description,\cite{Dal95,Dup90}
as can be seen in Figs.~\ref{fig3}-\ref{fig6} (the blurrier the
bubble-helium interface, the larger the surface thickness).
The zero temperature OT result, about 6 \AA{},
is in agreement with the experimental findings.\cite{Har98,Pen00}

\textit{c}. The density oscillations traveling through the liquid. This is the
point at which the differences between the two functionals become more
apparent. The density waves produced by the ST functional have much larger wavelengths
because this approach cannot sustain short wavelength inhomogeneities
due to the huge energy cost fromthe $|\nabla \rho(\bm{r})|^2$ 
surface energy term. The OT functional has not such a term and is free
from this drawback. Roughly speaking, the short wavelength waves arising in the
OT approach are smeared out in a sort of big {\it tsunami} in the ST case,
see for instance the panels corresponding to $t=5$ and 10~ps in 
Fig. \ref{fig6}.
It is worth emphasizing that the wave interference pattern found in the OT 
description is not an artifact produced by waves bouncing back
from the box boundaries, as those are already washed out by the damping term. It
arises from the interference of waves produced at different points on the 
e-bubble surface.

To quantitatively study the nature of the waves emitted
during the bubble evolution, we perform a Fourier analysis of the density
profile along the symmetry axis, restricting it to the region
30~\AA{} $\leq z \leq$ 70~\AA{}, away from the bubble location to avoid
uncontroled effects arising from the bubble itself. The Fourier
transform of the
density fluctuation is shown in Fig.~\ref{fig8} at $t=8.5$~ps and $P=0$~bar.
While both functionals generate low-$q$ density waves in the
phonon region (see Fig.\ref{fig1}) the ST approach does not display
any structure, whereas in the OT approach one can identify two distinct peaks.
The higher one is located at $q \sim 0.8$~\AA$^{-1}$ near the maxon region, 
and the lower one is located at $q \sim 2.3$ \AA$^{-1}$ close to the roton minimum.

With these results on the evolution of the 1P e-bubble in mind, we can state
with some confidence that the ST description is accurate enough for describing 
the fate of the e-bubble, yielding the appropriate
final topology at a given pressure, and a more than
qualitative picture of its evolution.
The shape of the cavity, which is responsible for
most electron properties, is essentially the same in both ST and OT
descriptions. The different way of energy release associated
with their each kind of elementary excitation may yield
somewhat diverse evolutions at longer times, but it has little
relevance for the problems at hand. One should keep in mind 
however, that if the actual subject of the study are the elementary 
excitations of the bulk liquid, the use of the OT functional is then
unavoidable.

We also want to stress that computing the 1P e-bubble dynamics in real time for
the ST functional has allowed us to explicitly
check the adiabatic approximation in the electron evolution during
the time interval relevant for the e-bubble ``fission''.\cite{Mat10a}
We have computed the overlap between the time-evolving electron wave
function and
the instantaneous eigenstate $1\sigma^-$, and have found it to be
equal to unity at all times, meaning that the adiabatic approximation holds.
As we will discuss later on, this is not the case for the 2P e-bubble.

\subsubsection{Time-resolved absorption spectrum}

Within the OT functional plus adiabatic approximation scheme, we have studied the
excitation of 1P bubbles by photoabsorption either to the
$m=0$ component ($2\sigma^+$), or to the $m=\pm 1$ components ($1\pi^-$),
arising from the splitting of the originally spherical 1D 
state, see Fig. \ref{fig2}. In principle,
this can be measured in a pump-probe experiment by which the e-bubble is
excited by two consecutive laser pulses. The delay set between these pulses
corresponds to the time interval between the excitation and the measurement, which 
is the same as the time defined in our calculations. The intensity of the
absorption lines is characterized here by their oscillator strength
calculated in the dipole approximation\cite{Wei78}

\begin{equation}
f_{ab} = \frac{2m_{\mathrm{e}}}{3\hbar^2}(E_a-E_b)\big|\langle a | \bm{r} | b
\rangle\big|^2 \;\, . 
\label{eq8}
\end{equation}
We recall that this oscillator strength fulfills the sum rule $\sum_a
f_{ab}=1$, but is generally not positive-definite. If the initial state is not
the ground state, a partial sum may be greater than unity.

Starting from the 1P electron state $1\sigma^-$ ($m=0$), the two possible
photoexcitation transitions are $1\sigma^-\rightarrow 2\sigma^+$ and
$1\sigma^-\rightarrow 1\pi^-$, see Fig. \ref{fig2}.
The specularly asymmetric states have a nodal
point on the $z=0$ symmetry plane, implying that they are rather insensitive to
the presence of helium in that plane. Therefore, the transition energy
for excitations between two asymmetric states should not
depend much on whether the bubble has split or not. Contrarily,
the specularly symmetric states do not have such a nodal point, and so are more
sensitive to splitting. The lowest-lying transition connecting specularly asymmetric
with  specularly symmetric states may thus
probe the topology of the bubble, since the absorption
energy for this transition should increase by a sizeable amount when the bubble
splits. The level structure at the right part of Fig. \ref{fig2}
may help understanding these issues.

The time-resolved absorption energies and oscillator strengths  
of the evolving 1P bubble for $P=0.5$ and 1 bar
are presented in Figs.~\ref{fig9} and \ref{fig10}.
As shown by our calculations, the 1P bubble does not split
for $P=0.5$~bar, but it does for $P=1$~bar at $t\simeq 170$~ps.
The bubble splitting yields a clear signature in the energies and the
oscillator strengths:
the evolution of the transition energies is similar
for $P=0.5$ and $P=1$ bar before the splitting,  but when the bubble
splits at $P=1$ bar, the $1\sigma^-\rightarrow 2\sigma^+$ energy
rapidly increases by $\sim 70$ meV, becoming comparable to the
$1\sigma^-\rightarrow 1\pi^-$ energy. This is a consequence of the
change in the bubble topology, which makes the final symmetric and antisymmetric
states nearly degenerate.

A conspicuous pattern also appears in the evolution of the oscillator strength.
The strength for the specularly asymmetric transition 
$1\sigma^-\rightarrow 1\pi^-$ remains nearly constant at $f \sim 0.65$,
whereas the strength for the specularly symmetric transition
$1\sigma^-\rightarrow 2\sigma^+$ oscillates  when the
bubble is simply connected but falls down to $f \sim 0.32$ when the bubble splits.
This is again a consequence of the near degeneracy of symmetric
and asymmetric states in the split-bubble regime.
The oscillator strength for the antisymmetric transition is a factor of two
larger than that of the symmetric transition in the split-bubble regime
because the final state $1\pi^-$ is twofold degenerate.\cite{note}

We thus conclude that time-resolved absorption energies are of practical 
interest because they bring rich information on the bubble shape and
can be determined in experiments. This may shed light
on the longstanding question about whether 1P e-bubbles under pressure do
really ``fission'' into two baby bubbles as our calculations indicate,
and how the electron wavefunction collapses into one of them, without 
violating the quantum measurement axiom.

\subsection{2P e-bubble dynamics: the breakdown of adiabaticity}

Our previous analysis of the dynamics of the 1P e-bubble has shown that one
does not need to use the accurate OT functional to describe
this process. The much simpler ST approach already yields a fair
description. This is particularly useful when we
move to the study of the 2P e-bubble dynamics. An attempt to simulate
this evolution has been made within the OT approach and the adiabatic
approximation.\cite{Mat10a} This
could only be performed for a few picoseconds, as it was shown that the adiabatic 
approximation fails at $t\simeq 7.4$ ps. This failure is due to the approaching 
of the  $3\sigma^-$ (arising from the spherical 2P level) and $2\sigma^-$
(arising from the spherical 1F level) energy levels.
By using the efficient real-time ST scheme,
we now relax the adiabatic approximation, following the evolution of the 2P e-bubble
for several hundreds of picoseconds.
We keep refering to this bubble as a ``2P e-bubble'', but should have in mind that
once the adiabatic approximation breaks down, the electron is no longer in
the original eigenstate. Generally, it is in a superposition of states that
have the same quantum numbers  as the initial 
state, meaning that it can be in any superposition of $\sigma^-$ states.

The 2P bubble evolution is shown in Fig.~\ref{fig11}.
For the first 100~ps, the shape evolution of the 2P bubble is similar to
that of the 1P bubble, as it expands along the symmetry $z$-axis while
its waist shrinks in the perpendicular plane.
From this point on, the bubble oscillates back and forth 
in a kind of four-lobe shapes quite different from those seen in the 1P bubble.
We attribute these conspicuous shape variations
to the breaking down of the adiabatic approximation as the electron moves 
from a eigenstate to a nontrivial superposition of those compatible with the
symmetries of the system. After evolving for $\sim$ 325~ps,
the 2P bubble splits into two baby bubbles.

We present in Fig.~\ref{fig12}(a) the evolution of the 
{\it instantaneous} eigenenergies of the first $\sigma^-$ states. 
As can be seen in panel (b),
the $2\sigma^-$ and $3\sigma^-$ states nearly meet at $t\simeq 7.4$~ps.
In agreement with some well-known results from basic quantum mechanics,\cite{Tel37}
we have found that this situation corresponds to an avoided crossing.
Panel (c) shows the overlap of the evolving electron wavefunction with
the relevant instantaneous eigenstates.
The electron is initially in a $3\sigma^-$ state (the overlap is unity), but at the point of avoided
crossing the adiabaticity is lost:
the electron state is a superposition of the $2\sigma^-$  ($\sim$ 80\%)
and $3\sigma^-$ ($\sim$ 20\%) states.

We have also found a time interval between $\sim$ 155 and $\sim$ 180~ps 
when the 2P bubble at $P=0$ bar has split but
the emerging baby bubbles do not go away. 
When this happens, the $n\sigma^-$ and $n\sigma^+$ states should be
degenerate. This is illustrated in panel (a) of Fig. \ref{fig12}
for the $n=1$ states. Notice from Fig. \ref{fig11} that in the
$50~{\rm ps} \lesssim  t \lesssim 100~{\rm ps}$ interval the bubble is
simply connected and
the apparent degeneracy displayed in Fig. {\ref{fig12}(c) is due
to the energy scale. The same thing happens around $t \sim 230$~ps.

\section{Summary}

We have thoroughly studied the dynamical evolution of 1P and 2P excited 
electron bubbles in superfluid $^4$He at zero temperature. To this end,
we have resorted to zero- and finite-range density functionals, 
establishing how reliable the former is by comparing its results with those obtained with the latter.

Although the results obtained for the 1P bubble evolution from these 
two functionals show some quantitative differences, especially for 
long-time evolutions, they are qualitatively equivalent. 
In particular, both lead to the conclusion that 1P bubbles ``fission'' 
at pressures above 1 bar. The ST functional result is of particular relevance, 
as it has been obtained by a real-time evolution, without assuming 
the adiabaticity of the process.
This confirms the previous results obtained using the finite-range
OT functional and the adiabatic approximation for much shorter
periods of time than in the present work.\cite{Mat10a}

Some indirect experimental evidence indicates a change in the
de-excitation behavior of the 1P e-bubble as pressure increases.\cite{Gri90,Mar04}
We have explored here the possibilities offered by the photoabsorption
spectrum of the 1P e-bubble to disclose whether such a bubble de-excites 
by ``fission'' or by a more conventional radiative decay, and have 
obtained the signatures that would help distinguish between both decay channels. 
Although far from trivial, a pump-probe experiment may detect a change 
in the absorption spectrum of the 1P bubble associated with
the appearance of the two baby bubble de-excitation channel.

Finally, we have studied the evolution of the 2P e-bubble in real-time within
the ST functional approach. We have dynamically
found that the adiabatic approximation does not hold at any positive
pressure confirming the results obtained within the OT plus 
adiabatic approximation approach.\cite{Mat10a} Negative pressures,
as those attained in cavitation experiments, have not been studied. The physical realization of
a relaxed, metastable 2P configuration is discarded.

\section*{Acknowledgments}

The authors wish to thank Humphrey Maris for helpful discussions. This work was
performed under Grants No. FIS2008-00421/FIS from DGI, Spain (FEDER), and
2009SGR1289 from Generalitat de Catalunya. D. Mateo has been supported by the
Spanish MEC-MICINN through the FPU fellowship program, Grant
No. AP2008-04343. D. Jin has been
supported by the National Science Foundation of the United States through Grant
No. DMR-0605355.

\pagebreak

\begin{figure}[f]
\centerline{\includegraphics[width=1.\linewidth,clip]{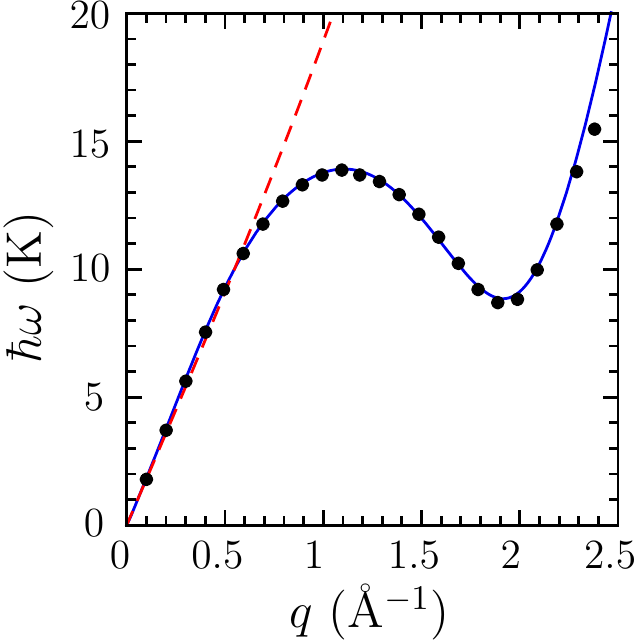}}
\caption{(Color online)
Dispersion relation of the elementary excitation in bulk liquid $^4$He at $T=0$.
Solid line: the OT finite-range functional results. Dashed
line: the ST zero-range functional results.
Dots: the experimental data from Ref.~\onlinecite{Don81}.
}
\label{fig1}
\end{figure}

\begin{figure}[f]
\centerline{\includegraphics[width=1.\linewidth,clip]{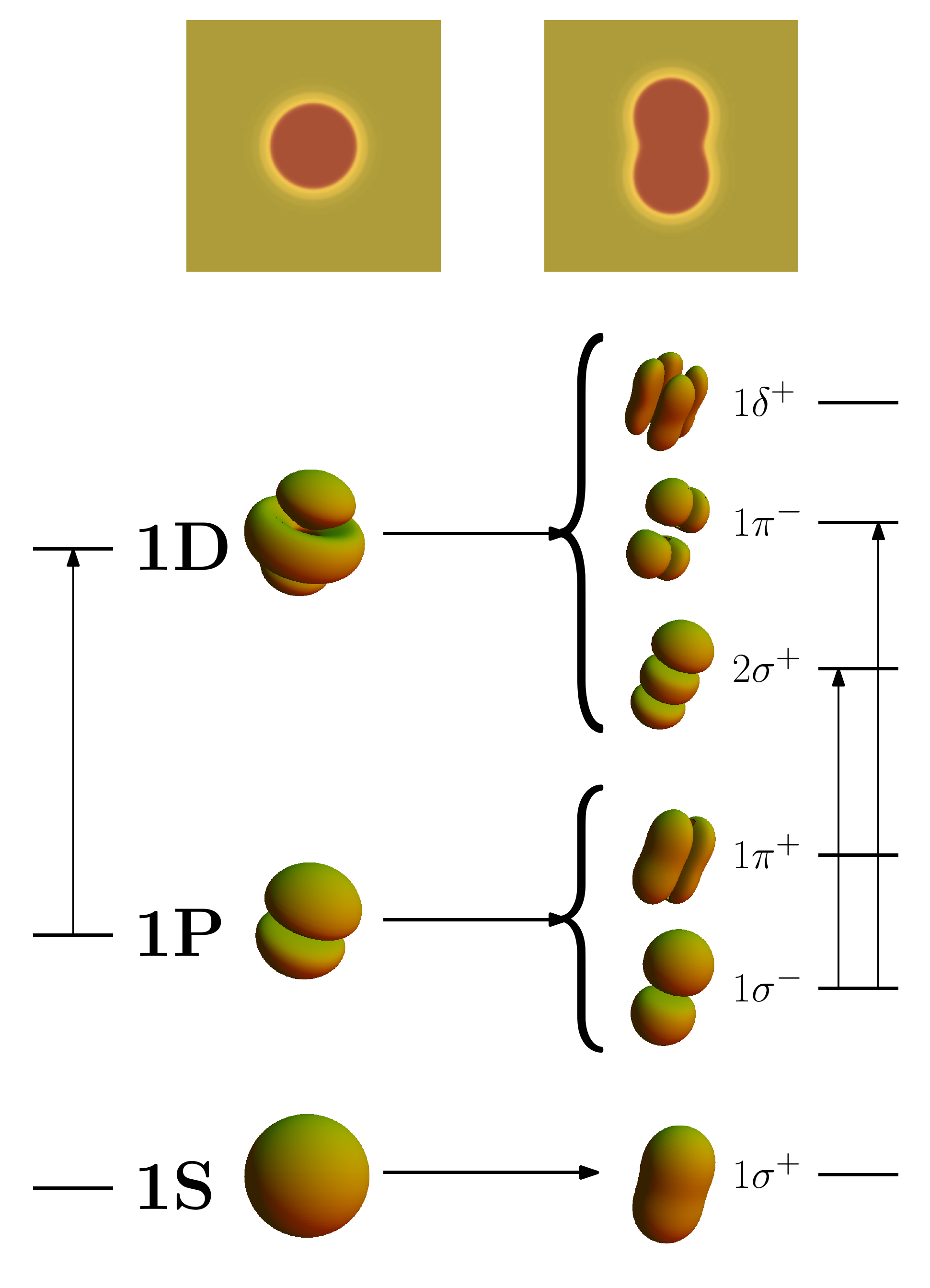}}
\caption{(Color online) 
Splitting of electronic levels along with a representation of their
probability densities once the spherical symmetry is broken.
The states in the spherical configuration (left) are labeled in the
standard $n$L way.
In the axially symmetric case (right) the label is $nl_z^s$, where 
$s=+ (-)$ stands for symmetric (antisymmetric) states under specular
reflection.
} 
\label{fig2}
\end{figure}

\begin{figure}[f]
\centerline{\includegraphics[width=1.\linewidth,clip]{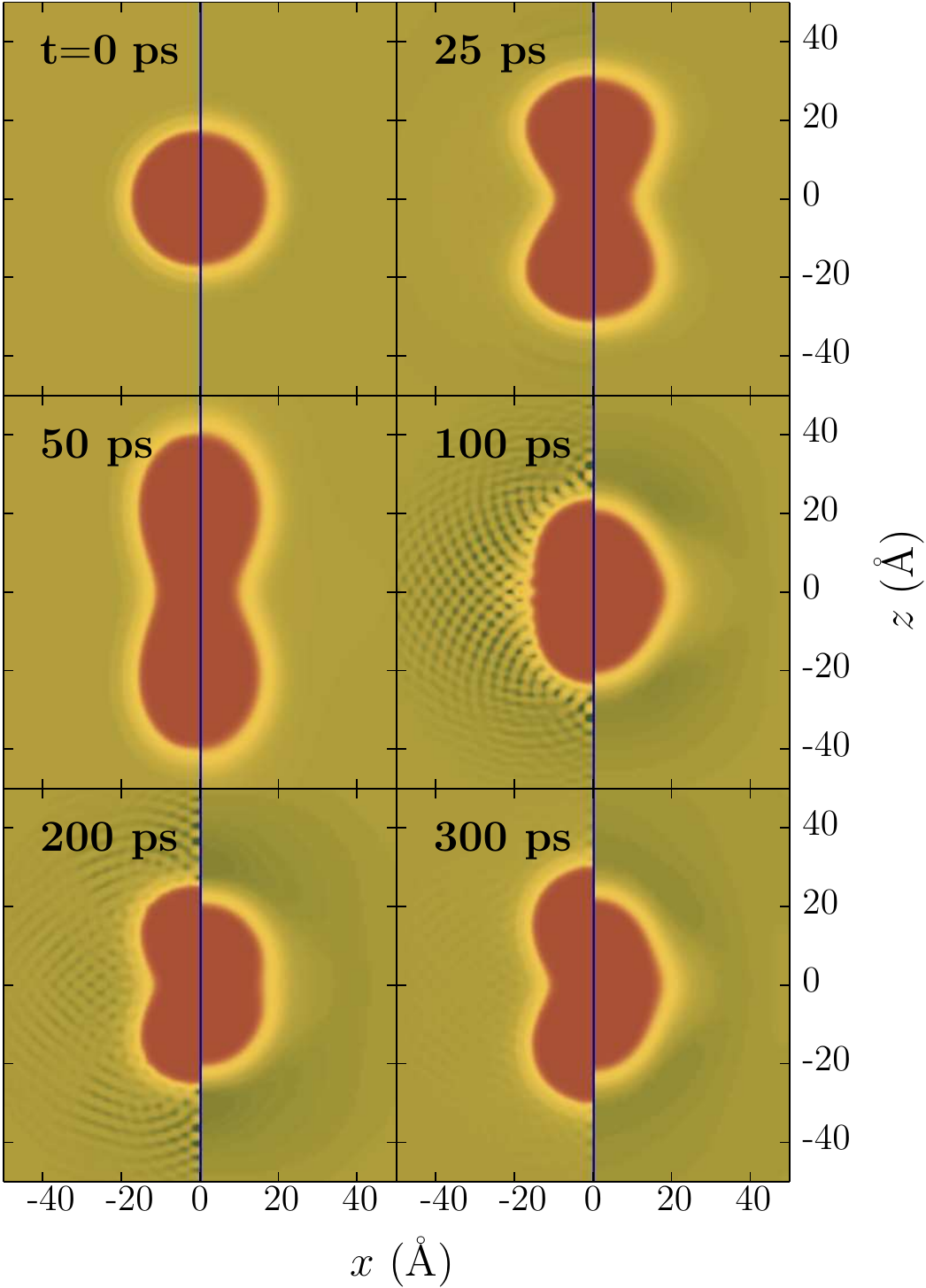}}
\caption{(Color online)
Evolution of the 1P e-bubble at $P=0$ bar. The left-hand side of each
panel shows the results for
the OT finite-range density functional plus adiabatic approximation for the electron.
The right-hand side part of each panel shows the results for the 
ST zero-range density functional plus real-time evolution for the electron.
}
\label{fig3}
\end{figure}

\begin{figure}[f]
\centerline{\includegraphics[width=1.\linewidth,clip]{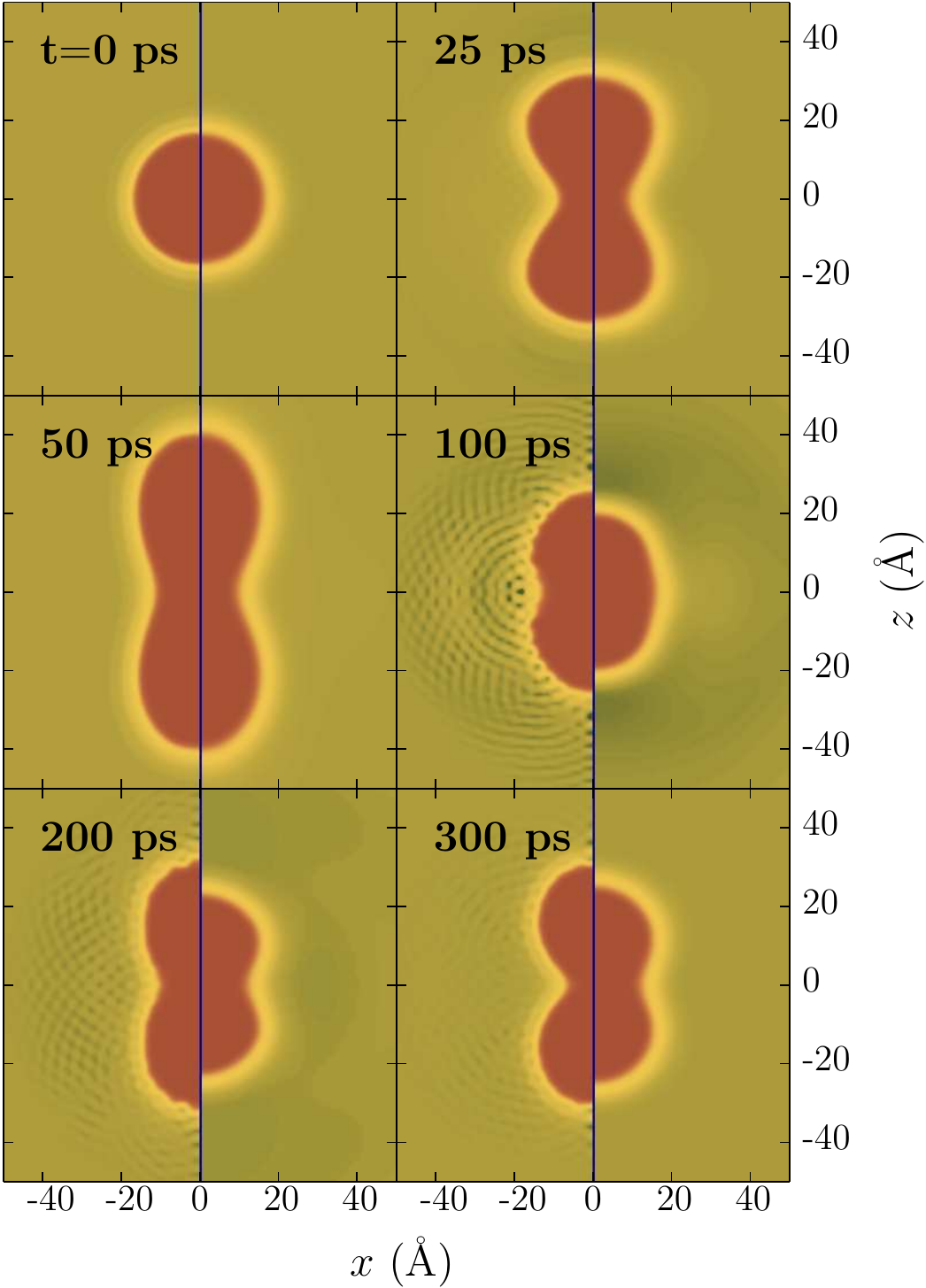}}
\caption{(Color online)
Same as Fig. \ref{fig3} at $P=0.5$ bar.
}
\label{fig4}
\end{figure}

\begin{figure}[f]
\centerline{\includegraphics[width=1.\linewidth,clip]{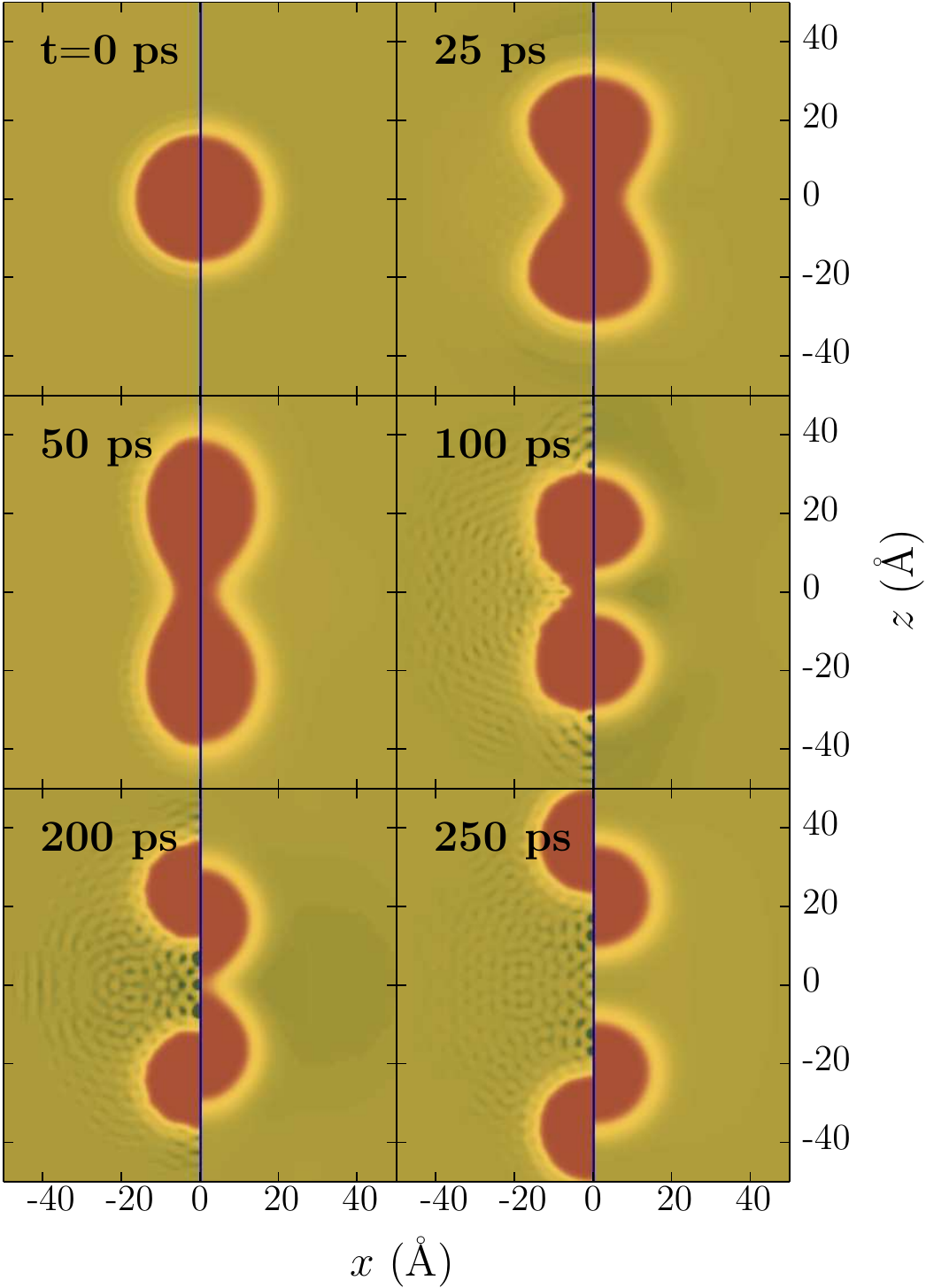}}
\caption{(Color online)
Same as Fig. \ref{fig3} at $P=1$ bar.
}
\label{fig5}
\end{figure}

\begin{figure}[f]
\centerline{\includegraphics[width=1.\linewidth,clip]{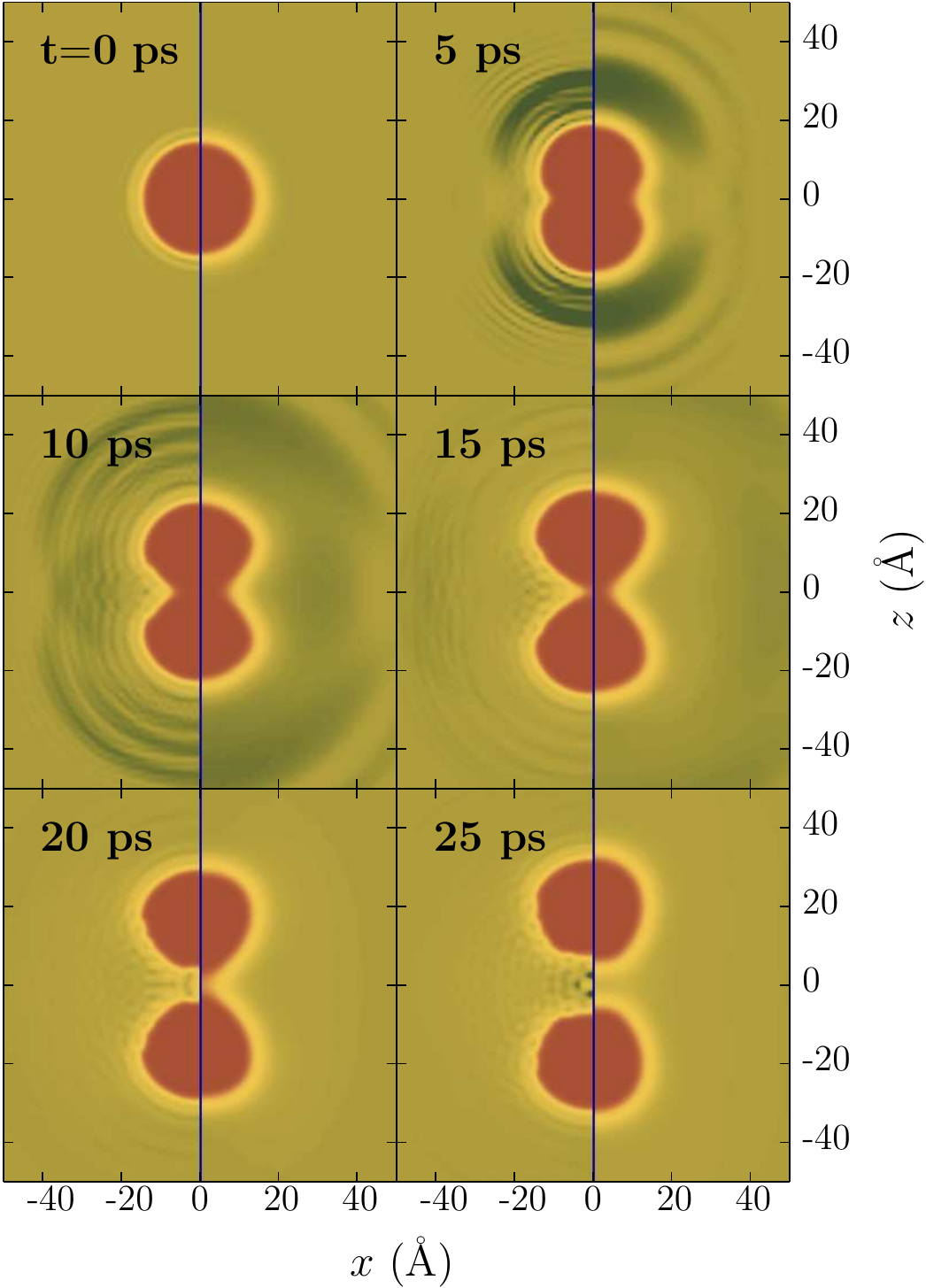}}
\caption{(Color online)
Same as Fig. \ref{fig3} at $P=5$ bar.
}
\label{fig6}
\end{figure}

\begin{figure}[f]
\centerline{\includegraphics[width=1.\linewidth,clip]{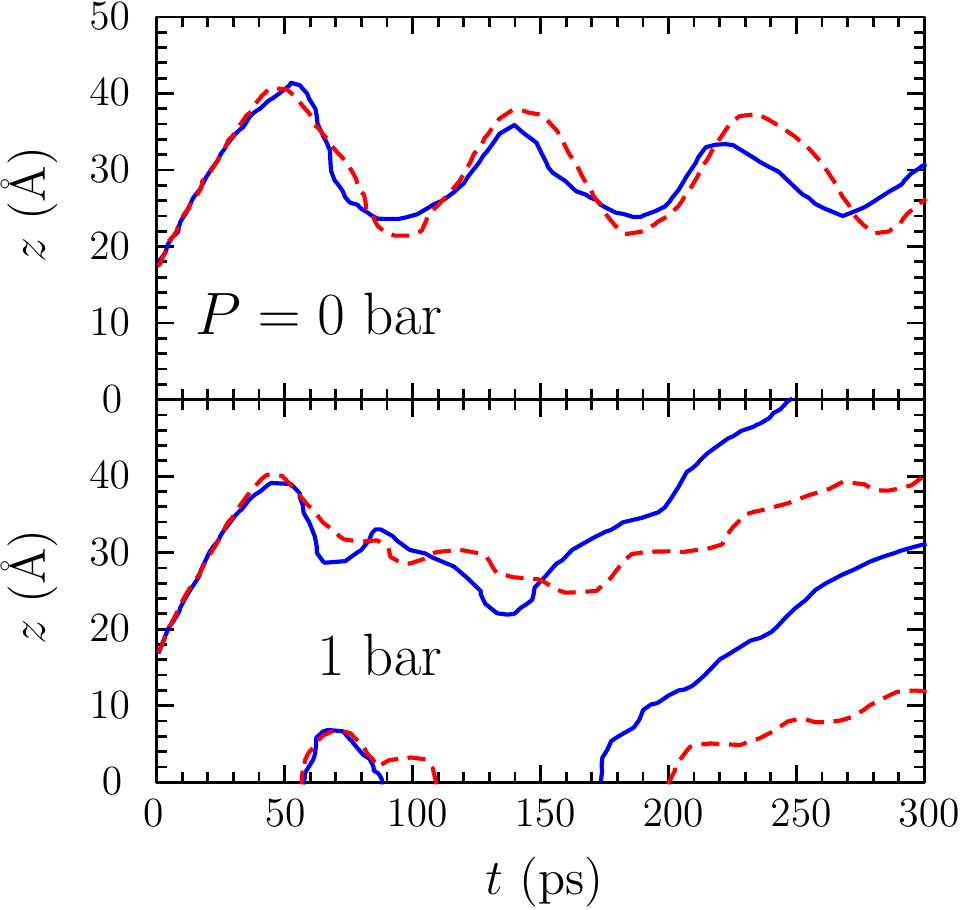}}
\caption{(Color online)
Evolution of the extent of the bubble along the $z$ axis at
$P=0$ and 1 bar. The
solid line is the OT finite-range result, and the dashed line is
the ST zero-range result.
}
\label{fig7}
\end{figure}

\begin{figure}[f]
\centerline{\includegraphics[width=1.\linewidth,clip]{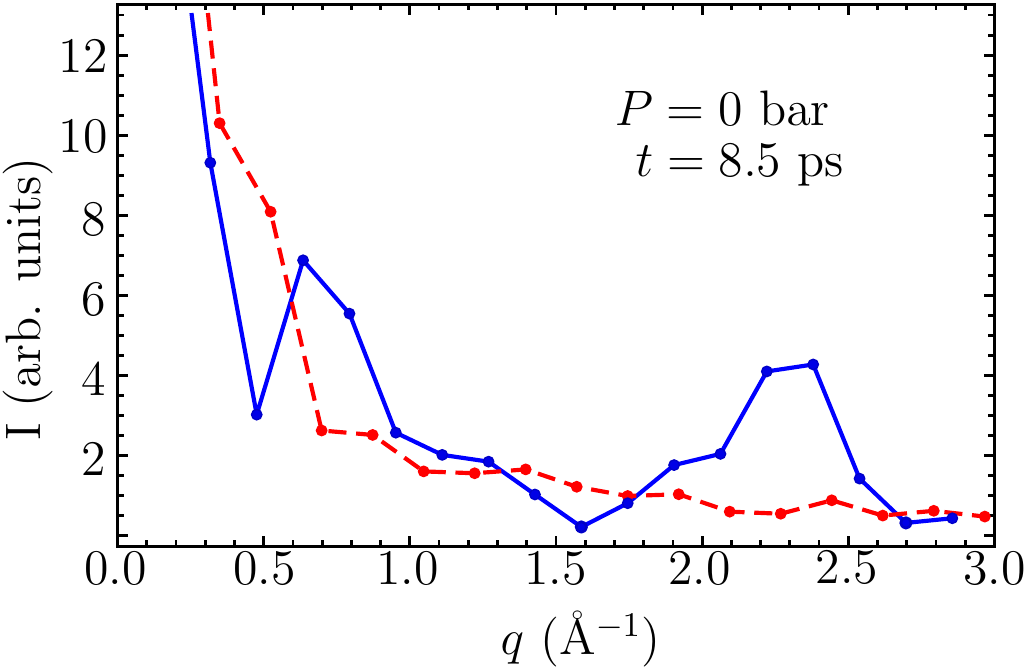}}
\caption{(Color online)
Fourier transform of the density fluctuation along the $z$-axis within
the region 30 \AA{} $\le z \le$ 70 \AA{} for the
expansion process of the 1P e-bubble at $P=0$ bar and  $t=8.5$ ps.
The solid line corresponds to the OT calculation, and the dashed line corresponds to the
ST functional calculation. The lines have been drawn as a guide to the eye.
}
\label{fig8}
\end{figure}

\begin{figure}[f]
\centerline{\includegraphics[width=1.\linewidth,clip]{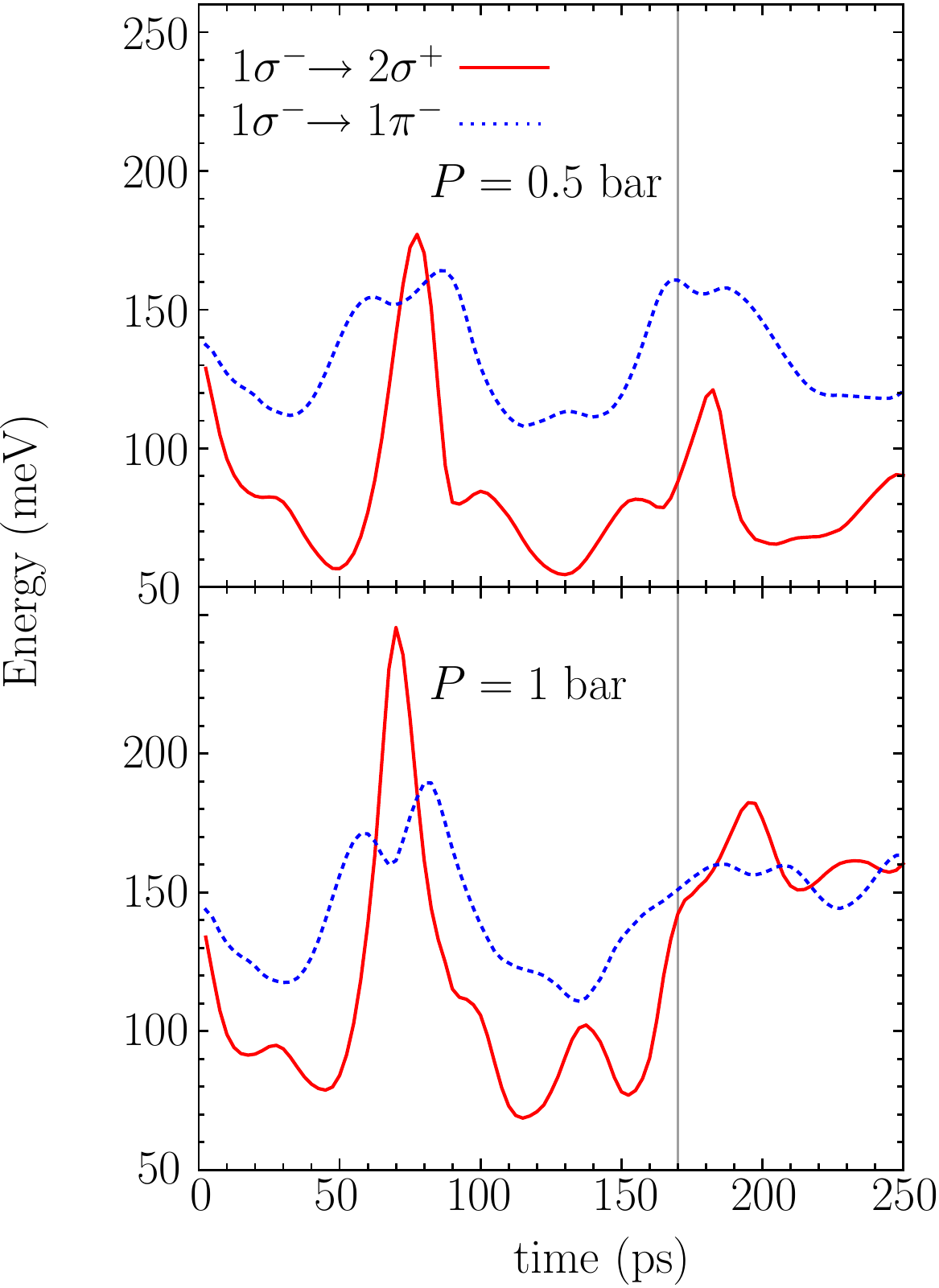}}
\caption{(Color online)
Time-resolved absorption energies at $P=0.5$ and 1 bar
for the 1P e-bubble evolution. The thin vertical line at $t=170$ ps
indicates the time at which the bubble splits in the $P=1$ bar
case.
}
\label{fig9}
\end{figure}

\begin{figure}[f]
\centerline{\includegraphics[width=1.\linewidth,clip]{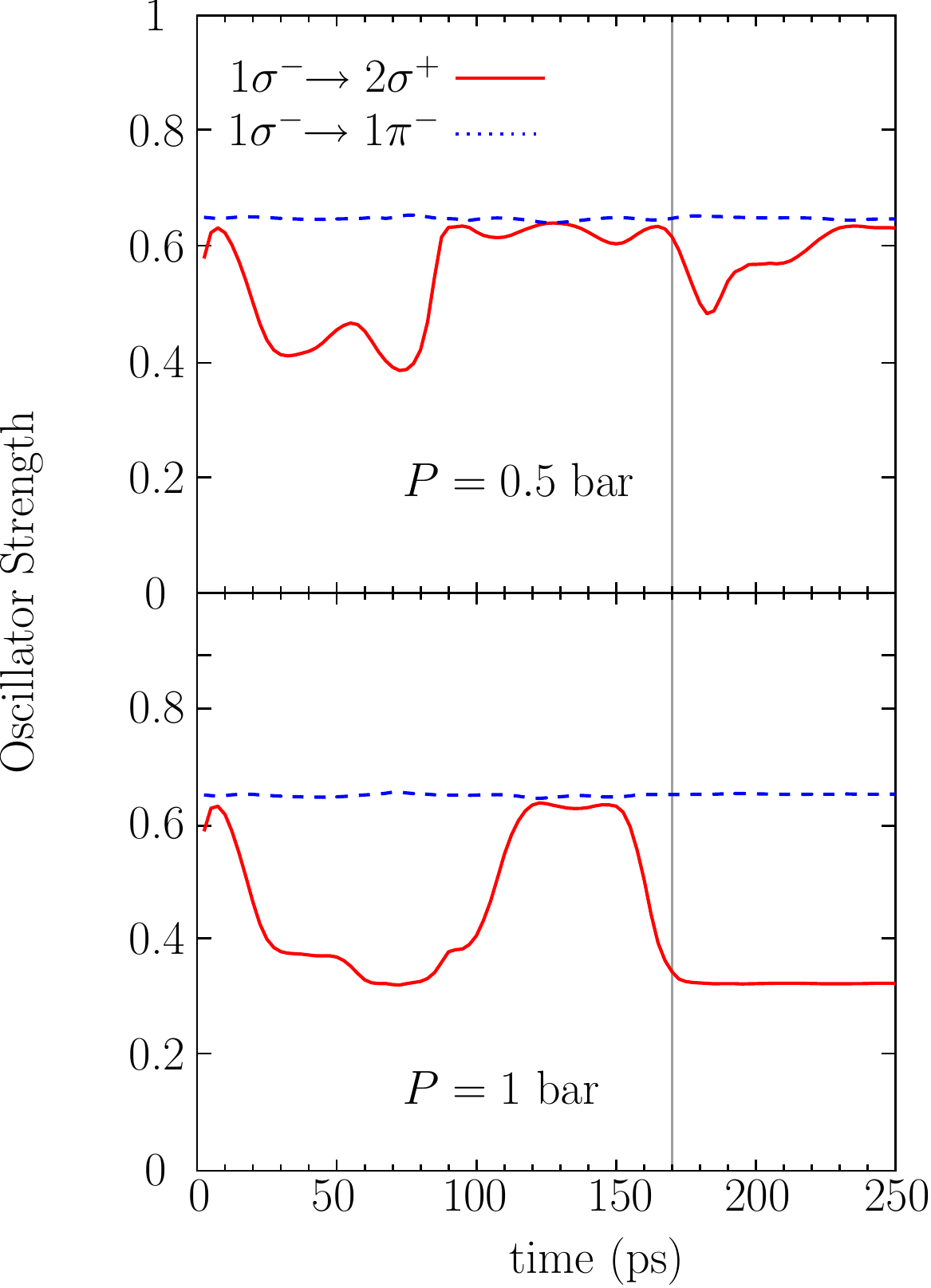}}
\caption{(Color online)
Time-resolved absorption oscillator strengths  at $P=0.5$ and 1 bar
for the 1P e-bubble evolution. The thin vertical line at $t=170$ ps
indicates the time at which the bubble splits at $P=1$ bar
}
\label{fig10}
\end{figure}

\begin{figure}[f]
\centerline{\includegraphics[width=1.\linewidth,clip]{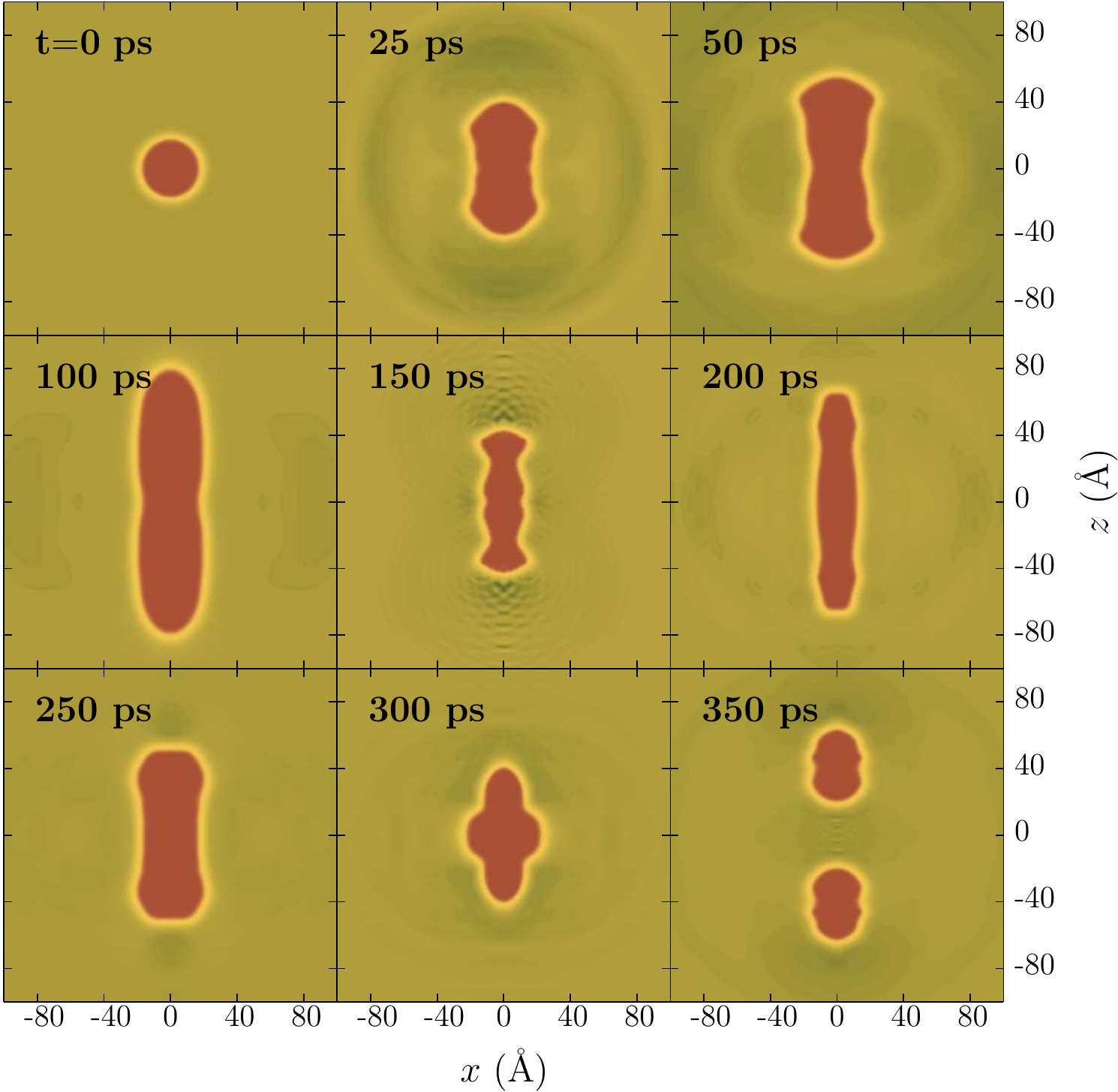}}
\caption{(Color online)
Evolution of the 2P e-bubble at $P=0$ bar using real-time dynamics and
the ST zero-range functional.
}
\label{fig11}
\end{figure}

\begin{figure}[f]
\centerline{\includegraphics[width=1.\linewidth,clip]{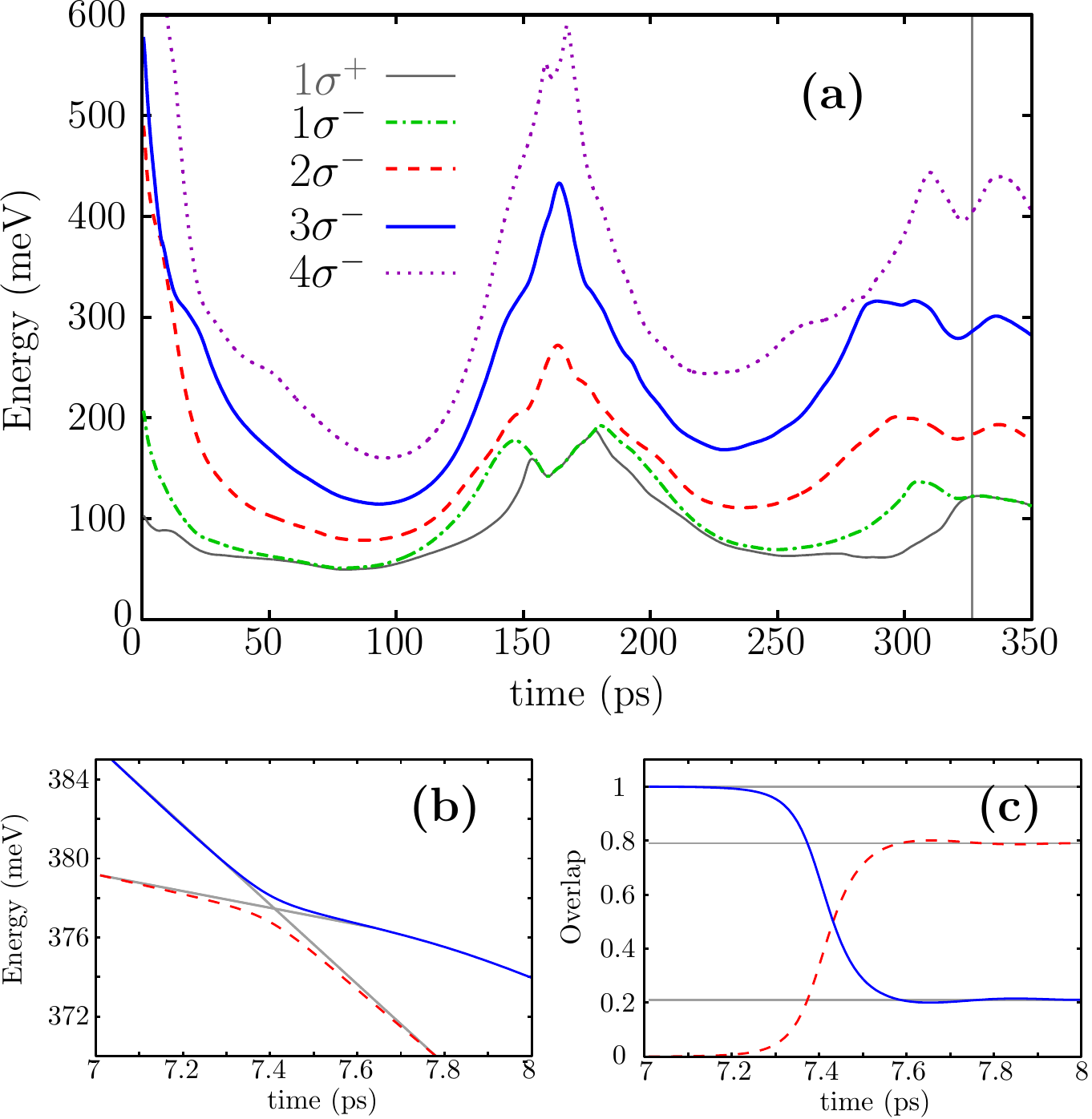}}
\caption{(Color online)
(a):
Lower-lying {\it instantaneous} $\sigma^-$ eigenstates 
together with the $1\sigma^+$ eigenstate of the
2P e-bubble at $P=0$ bar as a function of time.
The thin vertical line at $t=325$ ps 
indicates the time at which the bubble splits.
(b) Enlarged view of the region where the
$3 \sigma^-$ and $2 \sigma^-$ states repel each other.
(c) Overlap of the time-evolving
electron state onto the $3 \sigma^-$ (solid line)
and $2 \sigma^-$ eigenstates (dashed line), 
$|\langle \Phi(\bm{r}, t) |n\sigma^-\rangle|^2$.
The adiabatic approximation
fails if the value of this overlap varies in time.
}
\label{fig12}
\end{figure}

\end{document}